\documentclass[letterpaper,11pt]{article}
\usepackage{graphicx}
\usepackage[margin=1in]{geometry}
\usepackage[breaklinks,hidelinks]{hyperref}
\usepackage{amsthm}
\usepackage{amsmath}
\usepackage{amssymb}
\usepackage{accents}

\newtheorem{theorem}{Theorem}[]

\newtheorem{remark1}[theorem]{Remark}
\newenvironment{remark}{\begin{remark1} \rm}{\end{remark1}}

\title{Regression-aware decompositions}
\author{Mark Tygert}
\date{Facebook Artificial Intelligence Research}

\begin{document}

\maketitle

\begin{abstract}
Linear least-squares regression with a ``design'' matrix $A$
approximates a given matrix $B$ via minimization
of the spectral- or Frobenius-norm discrepancy $\|AX-B\|$
over every conformingly sized matrix $X$.
Another popular approximation is low-rank approximation
via principal component analysis (PCA)
--- which is essentially singular value decomposition (SVD) ---
or interpolative decomposition (ID).
Classically, PCA/SVD and ID operate solely
with the matrix $B$ being approximated,
not supervised by any auxiliary matrix $A$.
However, linear least-squares regression models
can inform the ID, yielding regression-aware ID.
As a bonus, this provides an interpretation as regression-aware PCA
for a kind of canonical correlation analysis between $A$ and $B$.
The regression-aware decompositions effectively enable supervision to inform
classical dimensionality reduction, which classically has been
totally unsupervised. The regression-aware decompositions reveal the structure
inherent in $B$ that is relevant to regression against $A$.
\end{abstract}

\section{Introduction}
\label{intro}

A common theme in multivariate statistics and data analysis is
detecting and exposing low-dimensional latent structure governing two sets
of vectors (each set could consist of realizations
of a vector-valued random variable, for example).
Widely used methodologies for this include linear least-squares regression
and the canonical correlation analysis (CCA) of~\cite{hotelling}
(\cite{hotelling} discusses all previously developed methods mentioned below).
Subsection~\ref{rapca} below combines the advantages
of both principal component analysis (PCA) and linear least-squares regression,
leveraging a single data set's intrinsic low-rank structure
as well as low-rank structure in the data's interaction with another data set;
this combination, ``regression-aware principal component analysis,''
amounts to a variant of CCA informed by the regression.

The core of the present paper is the construction in Subsection~\ref{raid}
of an analogous regression-aware interpolative decomposition,
which provides an efficient means of performing subset selection
for general linear models, especially in the simplified (while less canonical)
formulation of Subsection~\ref{raid2}.
The regression-aware interpolative decomposition selects some columns
of a given matrix $B$, then constructs numerically stable
(multi)linear interpolation from corresponding least-squares solutions
to the least-squares solutions $X = A^\dagger B$ minimizing $\| AX - B \|$
for all columns of $B$ (here, $A$ is the design matrix in the regression,
$A^\dagger$ is the pseudoinverse of $A$,
and $\| AX-B \|$ is the spectral or Frobenius norm).

Section~\ref{numex} illustrates these methods
via several numerical experiments. The other sections set the stage:
Section~\ref{prelims} reviews pertinent prior mathematics.
Section~\ref{constructs} introduces the regression-aware decompositions.
More specifically, Subsection~\ref{notation} specifies notational conventions.
Subsection~\ref{ids} defines and summarizes facts
about interpolative decompositions.
Subsection~\ref{general} formulates a general construction.
Subsection~\ref{specific} specializes the general formulation
of Subsection~\ref{general} to the case of linear least-squares regression,
albeit simplistically.
Subsection~\ref{raid} then provides the most useful formulation.
Subsection~\ref{rapca} leverages Subsection~\ref{raid} to interpret
a kind of CCA as a regression-aware decomposition.
Subsections~\ref{raid2} and~\ref{rapca2} provide computationally simpler
alternatives to Subsections~\ref{raid} and~\ref{rapca}, respectively.
Subsections~\ref{potential}--\ref{motion} present five illustrative
numerical examples.

\section{Preliminaries}
\label{prelims}

This section sets notation (in Subsection~\ref{notation})
and reviews the interpolative decomposition (in Subsection~\ref{ids}),
both of which are used throughout the remainder of the paper.

\subsection{Notation}
\label{notation}

This subsection sets notational conventions used throughout the present paper.

All discussion pertains to matrices whose entries are real- or complex-valued.
For any matrix $A$, we denote the adjoint (conjugate transpose) by $A^*$
and the Moore-Penrose pseudoinverse by $A^\dagger$;
we use $\|A\|$ to denote the same norm throughout the paper,
either the spectral norm or the Frobenius norm
(unitary invariance of the norm will be important in Section~\ref{constructs}),
and we denote by $(A^* A)^{-1/2}$ the pseudoinverse
of the self-adjoint square root of $A^* A$.
Detailed definitions of all these are available in the exposition
of~\cite{golub-van_loan}. A proof that $X = A^\dagger B$ minimizes
$\| AX-B \|$ for any conformingly sized matrices $A$ and $B$ ---
for both the spectral norm and the Frobenius norm ---
is available, for example, in Appendix~B of~\cite{szlam-tulloch-tygert};
accordingly, we refer to $X = A^\dagger B$
as ``the'' minimizer of $\| AX-B \|$.
All decompositions will be accurate to a user-specified precision
$\epsilon > 0$.

\subsection{Interpolative decomposition}
\label{ids}

This subsection reviews the interpolative decomposition (ID).

The ID dates at least to~\cite{fekete};
however, modern applications owe much to~\cite{tyrtyshnikov}
and~\cite{goreinov-tyrtyshnikov}, among others
(this is also related to the CX decomposition
of~\cite{drineas-mahoney-muthukrishnan} and others,
though technically the CX decomposition omits the ID's requirement
for numerical stability).
The software and documentation of~\cite{martinsson-rokhlin-shkolnisky-tygert}
describe some common algorithms for computing IDs,
based on the contributions of~\cite{cheng-gimbutas-martinsson-rokhlin}
and~\cite{gu-eisenstat}, which prove the following.

\begin{theorem}
\label{theorem}
Suppose that $m$ and $n$ are positive integers, and $B$
is an $m \times n$ matrix.

Then, for any positive integer $k$ with $k \le m$ and $k \le n$,
there exist a $k \times n$ matrix $P$ and an $m \times k$ matrix $C$
whose columns constitute a subset of the columns of $B$, such that
\begin{enumerate}
\item some subset of the columns of $P$ makes up the $k \times k$ identity
matrix,
\item no entry of $P$ has an absolute value greater than 1,
\item the spectral norm of $P$ is at most $\sqrt{k(n-k)+1}$,
\item the least (that is, the $k$th greatest) singular value of $P$
is at least 1,
\item $B = CP$ when $k = m$ or $k = n$, and
\item when $k < m$ and $k < n$,
\begin{equation}
\|B-CP\|_2 \le \sqrt{k(n-k)+1} \; \sigma_{k+1},
\end{equation}
where $\|B-CP\|_2$ is the spectral norm of the difference $B-CP$,
and $\sigma_{k+1}$ is the $(k+1)$th greatest singular value of $B$
(also, $\sigma_{k+1}$ is the spectral norm $\|B-\tilde{B}\|_2$
minimized over every $\tilde{B}$ whose rank is at most $k$).
\end{enumerate}
\end{theorem}

We often select $k$ in the theorem so that $\|B-CP\|_2$ is at most
some specified precision $\epsilon$. We say that $C$ collects together
a subset of the columns of $B$ and that $P$ is an interpolation matrix,
expressing to precision $\epsilon$ each column of $B$ as a linear combination
of the subset collected together into $C$. The factorization into the product
of $C$ and $P$ is known as an interpolative decomposition (ID). Properties 1--4
of Theorem~\ref{theorem} ensure that the ID is numerically stable.

Existing algorithms for computing $C$ and $P$ in Theorem~\ref{theorem} are
computationally expensive, so normally we instead require only that $C$ and $P$
satisfy the following weaker set of conditions:
\begin{enumerate}
\item some subset of the columns of $P$ makes up the $k \times k$ identity
matrix,
\item no entry of $P$ has an absolute value greater than 2,
\item the spectral norm of $P$ is at most $\sqrt{4k(n-k)+1}$,
\item the least (that is, the $k$th greatest) singular value of $P$
is at least 1,
\item $B = CP$ when $k = m$ or $k = n$, and
\item when $k < m$ and $k < n$,
\begin{equation}
\label{optimality}
\|B-CP\|_2 \le \sqrt{4k(n-k)+1} \; \sigma_{k+1},
\end{equation}
where $\|B-CP\|_2$ is the spectral norm of the difference $B-CP$,
and $\sigma_{k+1}$ is the $(k+1)$th greatest singular value of $B$
(also, $\sigma_{k+1}$ is the spectral norm $\|B-\tilde{B}\|_2$
minimized over every $\tilde{B}$ whose rank is at most $k$).
\end{enumerate}
For most purposes, the weaker conditions are essentially as useful
as those in Theorem~\ref{theorem}; moreover, there are many highly effective
algorithms for computing an ID which satisfies these.

\section{Mathematical constructions}
\label{constructs}

This section develops mathematical theory for regression-aware decompositions,
starting with the interpolative decomposition (ID)
in Subsections~\ref{general}--\ref{raid},
progressing to the singular value decomposition (SVD)
in Subsection~\ref{rapca}, and then simplifying
the requisite numerical computations
in Subsections~\ref{raid2} and~\ref{rapca2}.

\subsection{An ID with an auxiliary matrix}
\label{general}

This subsection provides a general formulation,
of which the following two subsections are special cases.  

Given matrices $A$ and $B$ of sizes conforming for the product $AB$,
we can form an ID of $AB$, collecting together
a subset of the columns of $AB$ into a matrix $AC$, where $C$ collects together
a subset of the columns of $B$, together with an interpolation matrix $P$:
\begin{equation}
\label{initial}
\| AB - ACP \| \le \epsilon.
\end{equation}
Expressing~(\ref{initial}) as
\begin{equation}
\label{final}
\| A(B-CP) \| \le \epsilon,
\end{equation}
we may view this as interpolating stably and accurately to all columns of $B$
from the subset collected together in $C$,
provided that the accuracy of the interpolation is measured via the ``norm''
in~(\ref{final}) involving $A$,
\begin{equation}
\|D\|_A = \|AD\|
\end{equation}
for any matrix $D$ of size conforming for the product $AD$,
including $D = B-CP$.

\subsection{An ID for regression}
\label{specific}

This subsection constructs an ID that is informed
by linear least-squares regression, attaining high accuracy
when measuring errors directly on the least-squares solutions
(which is a terrible idea in the typical, numerically rank-deficient case
of interest for dimensionality reduction). The following subsection alters
the simplistic formulation of the present subsection,
instead measuring errors via the residuals of the least-squares fits.

Substituting the pseudoinverse $A^\dagger$ for $A$ in Subsection~\ref{general},
we obtain the following:
Given matrices $A$ and $B$ of sizes conforming for the product $A^\dagger B$,
we can form an ID of $A^\dagger B$, collecting together
a subset of the columns of $A^\dagger B$ into a matrix $A^\dagger C$,
where $C$ collects together a subset of the columns of $B$,
together with an interpolation matrix $P$:
\begin{equation}
\label{initial2}
\| A^\dagger B - A^\dagger CP \| \le \epsilon.
\end{equation}
Denoting by $X$ the minimizer of $\| AX-B \|$ given by $X = A^\dagger B$
and by $Y$ the minimizer of $\| AY-C \|$ given by $Y = A^\dagger C$,
we may express~(\ref{initial2}) as
\begin{equation}
\label{final2}
\| X - YP \| \le \epsilon.
\end{equation}
Thus, the selected columns of $B$ collected together into $C$ enable
accurate interpolation from the corresponding least-squares solutions
to the least-squares solutions for all columns of $B$.

\subsection{A regression-aware ID}
\label{raid}

This subsection constructs a decomposition which answers the question
of how a matrix $B$ looks under the general linear model
with a given design matrix $A$, that is, how $B$ looks under the regression
which minimizes $\|AX-B\|$. ``Looks'' means that the decomposition
provides a subset of the columns of $B$ such that the least-squares solutions
for the subset can stably and to high precision be (multi)linearly interpolated
to the least-squares solutions ($X$) for all columns of $B$,
at least when measuring accuracy via the residuals $\|AX-B\|$.
[Statisticians, beware: $X$ denotes the solution $X=A^\dagger B$
to the linear least-squares regression minimizing $\|AX-B\|$,
not the design matrix. The design matrix is $A$.]

Here, given a matrix $A$, we define
\begin{equation}
\label{sqrts}
S = (A^* A)^{-1/2} A^*;
\end{equation}
notice that
\begin{equation}
\label{projector}
A A^\dagger = S^* S.
\end{equation}
Substituting $S$ for $A$ in Subsection~\ref{general}, we obtain the following:
Given matrices $A$ and $B$ of sizes conforming for the product $SB$,
we can form an ID of $SB$, collecting together a subset of the columns
of $SB$ into a matrix $SC$, where $C$ collects together a subset of the columns
of $B$, together with an interpolation matrix $P$:
\begin{equation}
\label{initial3}
\| SB - SCP \| \le \epsilon.
\end{equation}
Denoting by $X$ the minimizer of $\| AX-B \|$ given by $X = A^\dagger B$
and by $Y$ the minimizer of $\| AY-C \|$ given by $Y = A^\dagger C$,
combining~(\ref{projector}), (\ref{initial3}), the unitary invariance
of the norm, and the fact that each singular value of $S$
defined in~(\ref{sqrts}) is either 1 or 0 yields that
\begin{equation}
\label{final3}
\| AX - AYP \|
= \| A A^\dagger [B-CP] \|
= \| S^* S [B-CP] \|
= \| SB - SCP \| \le \epsilon.
\end{equation}
Thus, the selected columns of $B$ collected together into $C$ enable
numerically stable interpolation from the corresponding least-squares solutions
to the least-squares solutions for all columns of $B$, to high precision,
when measuring accuracy via the residuals. Indeed, (\ref{final3}) yields that
\begin{equation}
\bigl| \|AX-B\| - \|AYP-B\| \bigr| \le \|(AX-B)-(AYP-B)\| \le \epsilon.
\end{equation}

\begin{remark}
The nonzero singular values and corresponding right singular vectors
of $(A^* A)^{-1/2} A^*$ and of the self-adjoint square root of $A A^\dagger$
are the same; note also that $A A^\dagger$ is the self-adjoint square root
of itself --- $A A^\dagger$ is an orthogonal projector. So,
constructing an ID of $A A^\dagger B$ could select the same columns of $B$
and produce the same interpolation matrix $P$ as the above procedure,
which constructs an ID of $(A^* A)^{-1/2} A^* B$.
However, using $(A^* A)^{-1/2} A^*$ is more efficient when $A$
is tall and skinny.
\end{remark}

\subsection{Regression-aware principal component analysis}
\label{rapca}

The singular value decomposition (SVD) provides an alternative to using IDs.
Given matrices $A$ and $B$ of sizes conforming for the product $A^* B$,
the SVD of $(A^* A)^{-1/2} A^* B$ provides a kind of regression-aware
principal component analysis (PCA), as PCA and SVD are more or less the same.
This is basically the canonical correlation analysis (CCA) of~\cite{hotelling},
though CCA usually involves whitening $B$ to $B (B^* B)^{-1/2}$
prior to taking the SVD: the most popular formulation of CCA forms the SVD
of $(A^* A)^{-1/2} A^* B (B^* B)^{-1/2}$.
That said, the SVD of $(A^* A)^{-1/2} A^* B$ has the interpretation developed
in the previous subsection as a regression-aware PCA, even without whitening.

In detail, defining $S$ via~(\ref{sqrts}), we can form a low-rank approximation
of $SB$ with matrices $U$, $\Sigma$, and $V$ such that
\begin{equation}
\label{svd}
\| SB - U \Sigma V^* \| \le \epsilon,
\end{equation}
where the columns of $U$ are orthonormal, as are the columns of $V$,
the entries of $\Sigma$ are all nonnegative
and are zero off the main diagonal,
and the column span of $U$ lies in the column span of $S$.
Denoting by $X$ the minimizer of $\| AX-B \|$ given by $X = A^\dagger B$,
combining~(\ref{projector}), (\ref{svd}), the unitary invariance of the norm,
and the fact that each singular value of $S$ defined in~(\ref{sqrts})
is either 1 or 0 yields that
\begin{equation}
\label{finalsvd}
\| AX - S^* U \Sigma V^* \|
= \| A A^\dagger B - S^* U \Sigma V^* \|
= \| S^* S B - S^* U \Sigma V^* \|
= \| SB - U \Sigma V^* \| \le \epsilon.
\end{equation}
In particular, combining~(\ref{sqrts}) and (\ref{finalsvd}) yields that
\begin{equation}
\bigl| \|AX-B\| - \| A T \Sigma V^* - B \| \bigr|
= \bigl| \|AX-B\| - \| S^* U \Sigma V^* - B \| \bigr|
\le \| (AX-B) - (S^* U \Sigma V^* - B) \| \le \epsilon,
\end{equation}
where
\begin{equation}
T = (A^* A)^{-1/2} U.
\end{equation}
Thus, the reduced-rank representation $T \Sigma V^*$ permits reconstruction
of $B$ effectively as accurately as $X = A^\dagger B$,
the best possible minimizer of $\| AX-B \|$.
Admittedly, the interpretation here is not as satisfying as that
in Subsection~\ref{raid}, but is clearly strongly related just the same.
Singular vectors are linear combinations of the original vectors,
whereas the columns selected in Subsection~\ref{raid} are simply a subset
of the original vectors.

\subsection{Simpler computations}
\label{raid2}

This subsection provides a computationally simpler (albeit less natural)
version of Subsection~\ref{raid}.

Here, given a matrix $A$, we form a pivoted QR decomposition
\begin{equation}
\label{QR}
A = Q R \Pi,
\end{equation}
where the columns of $Q$ are orthonormal, $\Pi$ is a permutation matrix,
and $R$ is an upper-triangular (or upper-trapezoidal) matrix whose entries
on the main diagonal are all nonzero; notice that
\begin{equation}
\label{projectora}
A A^\dagger = Q Q^*.
\end{equation}
Substituting $Q^*$ for $A$ in Subsection~\ref{general},
we obtain the following:
Given matrices $A$ and $B$ of sizes conforming for the product $Q^* B$,
we can form an ID of $Q^* B$, collecting together a subset of the columns
of $Q^* B$ into a matrix $Q^* C$, where $C$ collects together a subset
of the columns of $B$, together with an interpolation matrix $P$:
\begin{equation}
\label{initiala}
\| Q^* B - Q^* C P \| \le \epsilon.
\end{equation}
Denoting by $X$ the minimizer of $\| AX-B \|$ given by $X = A^\dagger B$
and by $Y$ the minimizer of $\| AY-C \|$ given by $Y = A^\dagger C$,
combining~(\ref{projectora}), (\ref{initiala}), the unitary invariance
of the norm, and the fact that the columns of $Q$ from~(\ref{QR})
are orthonormal yields that
\begin{equation}
\label{finala}
\| AX - AYP \|
= \| A A^\dagger [B-CP] \|
= \| Q Q^* [B-CP] \|
= \| Q^* B - Q^* C P \| \le \epsilon.
\end{equation}
Thus, the selected columns of $B$ collected together into $C$ enable
numerically stable interpolation from the corresponding least-squares solutions
to the least-squares solutions for all columns of $B$, to high precision,
when measuring accuracy via the residuals. Indeed, (\ref{finala}) yields that
\begin{equation}
\bigl| \|AX-B\| - \|AYP-B\| \bigr| \le \|(AX-B)-(AYP-B)\| \le \epsilon.
\end{equation}

\subsection{Another way to regression-aware PCA}
\label{rapca2}

This subsection provides a computationally simpler (albeit less natural)
version of Subsection~\ref{rapca}.

Here, given a matrix $A$, we form a pivoted QR decomposition
\begin{equation}
\label{QR2}
A = Q R \Pi,
\end{equation}
where the columns of $Q$ are orthonormal, $\Pi$ is a permutation matrix,
and $R$ is an upper-triangular (or upper-trapezoidal) matrix whose entries
on the main diagonal are all nonzero.
We can form a low-rank approximation of $Q^* B$ with matrices
$U$, $\Sigma$, and $V$ such that
\begin{equation}
\label{svda}
\| Q^* B - U \Sigma V^* \| \le \epsilon,
\end{equation}
where the columns of $U$ are orthonormal, as are the columns of $V$,
the entries of $\Sigma$ are all nonnegative
and are zero off the main diagonal,
and the column span of $U$ lies in the column span of $Q^*$.
Denoting by $X$ the minimizer of $\| AX-B \|$ given by $X = A^\dagger B$,
combining~(\ref{projectora}), (\ref{svda}), the unitary invariance of the norm,
and the fact that the columns of $Q$ from~(\ref{QR2}) are orthonormal
yields that
\begin{equation}
\label{finalsvda}
\| AX - Q U \Sigma V^* \|
= \| A A^\dagger B - Q U \Sigma V^* \|
= \| Q Q^* B - Q U \Sigma V^* \|
= \| Q^* B - U \Sigma V^* \| \le \epsilon.
\end{equation}
In particular, combining~(\ref{QR2}) and (\ref{finalsvda}) yields that
\begin{equation}
\bigl| \|AX-B\| - \| A T \Sigma V^* - B \| \bigr|
= \bigl| \|AX-B\| - \| Q U \Sigma V^* - B \| \bigr|
\le \| (AX-B) - (Q U \Sigma V^* - B) \| \le \epsilon,
\end{equation}
where
\begin{equation}
T = \Pi^{-1} R^\dagger U.
\end{equation}
Thus, the reduced-rank representation $T \Sigma V^*$ permits reconstruction
of $B$ effectively as accurately as $X = A^\dagger B$,
the best possible minimizer of $\| AX-B \|$.
Since the columns of $QU$ are orthonormal
($(QU)^* (QU)$ is the identity matrix),
the singular values of $A T \Sigma V^* = Q U \Sigma V^*$
are the diagonal entries of $\Sigma$.

\section{Numerical examples}
\label{numex}

This section discusses several illustrative examples.
The section first presents two examples with synthetic data,
in Subsections~\ref{potential} and~\ref{synthseries},
then considers real data, in Subsections~\ref{ld}--\ref{motion}.
Software for running the examples is available at
\url{http://tygert.com/rad.tar.gz}

Several of the examples display biplots,
in the rightmost halves of Figures~\ref{synthplot}--\ref{g60};
regarding biplots, please consult~\cite{gabriel}
or~\cite{gower-lubbe-le_roux}.
The horizontal coordinates of the black circular dots in the biplots
are the leading ``scores'' --- the greatest singular value times
the entries of the corresponding left singular vector;
the vertical coordinates of the black circular dots
are the next leading scores --- the second greatest singular value times
the entries of the corresponding left singular vector.
The horizontal coordinates of the tips of the gray lines in the biplots
are the entries of the right singular vector
corresponding to the greatest singular value;
the vertical coordinates of the tips of the gray lines
are the entries of the right singular vector
corresponding to the second greatest singular value.

In Figures~\ref{synthplot}--\ref{g60}, $Q_A$ refers to a matrix
whose columns form an orthonormal basis for the column span of $A$,
and $Q_B$ refers to a matrix whose columns form an orthonormal basis
for the column span of $B$; the singular values $\sigma_k$ of $(Q_A)^* Q_B$
are those arising in the canonical correlation analysis (CCA)
between $A$ and $B$, whereas the singular values $\sigma_k$ of $(Q_A)^* B$
are those arising in the regression-aware principal component analysis (RAPCA)
of $B$ for $A$. The singular values for the RAPCA also determine
the spectral-norm accuracy of the regression-aware interpolative decomposition
(RAID), commensurate with formula~(\ref{optimality}).

\subsection{Potential theory}
\label{potential}

\begin{figure}
\centering
\includegraphics[width=.75\textwidth]{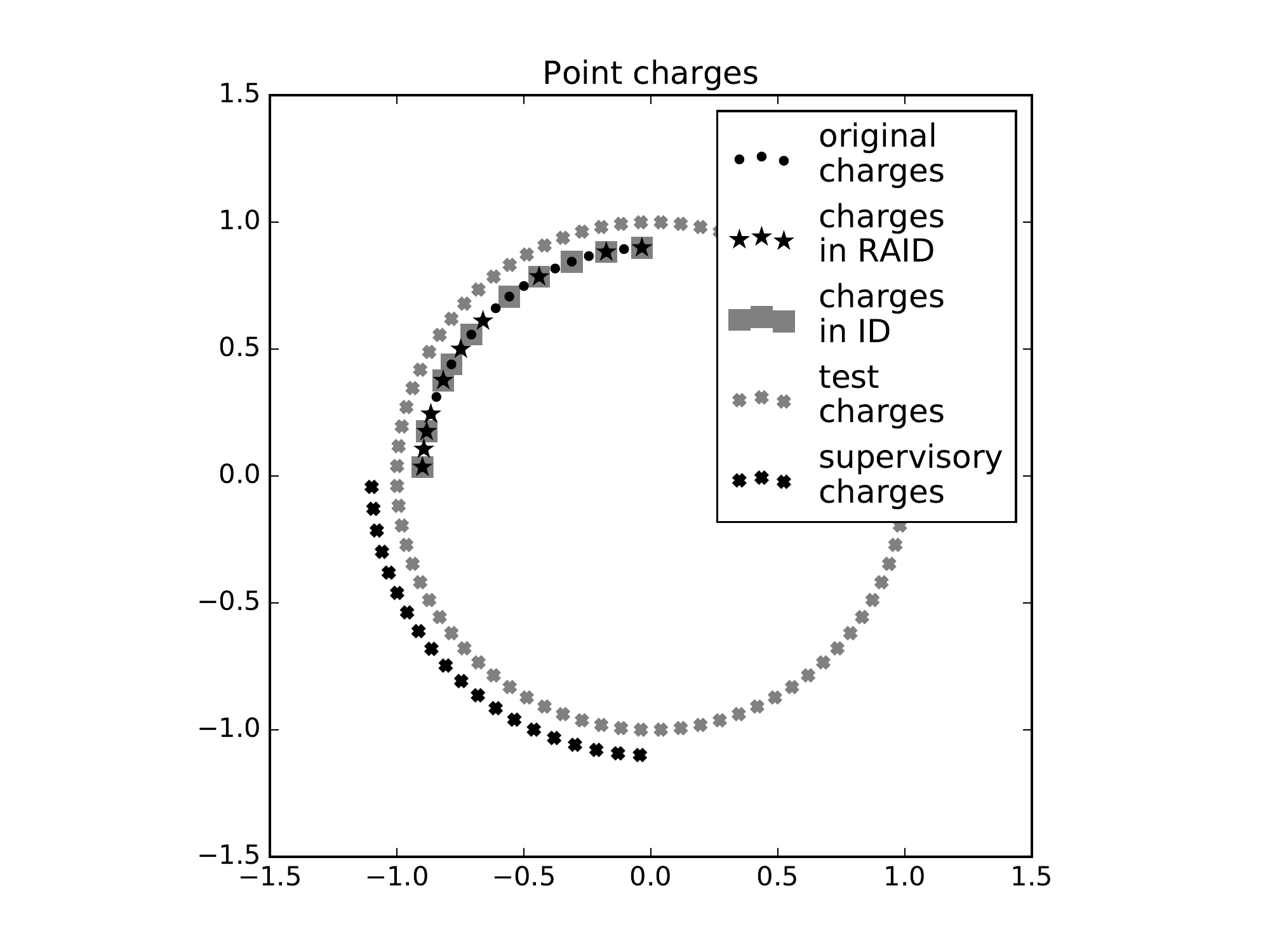}
\caption{Example from Subsection~\ref{potential}}
\label{charges}
\end{figure}

This subsection considers points on concentric circles
of radii 0.9, 1, and 1.1, as illustrated in Figure~\ref{charges}.
The interaction between a unit charge at point $p$ and a unit charge
at point $q$ is $\ln(\|p-q\|)$, the potential energy for the Laplace equation
in two dimensions, where $\|p-q\|$ denotes the Euclidean distance
between $p$ and $q$.
We are interested in the interactions of the test charges
in Figure~\ref{charges} with the original charges, but only those components
of the interactions that are representable by the interactions
of the test charges with the supervisory charges.
There are 80 test charges spread evenly around the circle of radius 1.
There are 20 original charges spread evenly around the top-left quadrant
of the circle of radius 0.9.
There are 20 supervisory charges spread evenly around the bottom-left quadrant
of the circle of radius 1.1.
The entries of an $80 \times 20$ matrix $B$ are the natural logarithms
of the distances between the test charges and the original charges,
normalized such that the spectral norm $\|B\|_2$ becomes 1.
The entries of an $80 \times 20$ matrix $A$ are the natural logarithms
of the distances between the test charges and the supervisory charges,
normalized by the same factor as $B$.
The ID of $B$ considered here selects 10 representative charges
from the original charges; the RAID of $B$ for $A$ selects a different set
of 10.
The spectral norm of the difference between $B$ and its reconstruction
from the ID is 0.016.
The spectral-norm error of the RAID is 0.25E--10;
the spectral-norm error is $\|AX-AYP\|_2$ from the left-hand side
of~(\ref{finala}).
(For reference, $\min_X \|AX-B\|_2 = 0.67$,
where $\|AX-B\|_2$ is the spectral norm of $AX-B$.)
Thus, the interactions between the test charges and the 10 charges selected
by the RAID are sufficient to capture to very high accuracy
the interactions between the test charges and all the original charges,
at least those components that are representable
by the interactions between the test charges and the supervisory charges.

\subsection{Synthetic time-series}
\label{synthseries}

This subsection analyzes a synthetic multivariate time-series,
specifically the matrix $C$ with $m =$ 10,000,000 rows and $n =$ 10 columns
constructed as follows:
We start with all entries being i.i.d.\ standard normal variates.
Then, we multiply the first 5 columns by 1,000,000,
and in each of the last 5 columns set all entries equal to the entry
in the last row (so that the last 5 columns are just multiples of each other).
Finally, we add to every entry 0.01 times the product
of the row and column indices (this is a rank-1 perturbation).
We let $A$ be the first block of $m-1$ rows
of $C$ and let $B$ be the last block of $m-1$ rows
(so $A$ includes the first row of $C$ but not the last,
whereas $B$ includes the last row of $C$ but not the first),
dividing each entry of $A$ and $B$ by the same factor such that
the spectral norm $\|B\|_2$ becomes 1.

The ID of $B$ considered here selects 4 representative columns
from the originals; the RAID of $B$ for $A$ selects a different set of 4.
Specifically, the ID ends up selecting columns 2--5, whereas the RAID
ends up selecting columns 1, 2, 5, and 10
--- the ID entirely misses the last 5 columns
(which were multiples of each other prior to adding the rank-1 perturbation),
whereas the RAID includes one of the second 5 columns (namely, the last).
The spectral norm of the difference between $B$ and its reconstruction
from the ID is 0.80.
The spectral-norm error of the RAID is 0.00039.
(For reference, $\min_X \|AX-B\|_2 = 0.79$,
where $\|AX-B\|_2$ is the spectral norm of $AX-B$.)
Thus, the 4 columns selected by the RAID are sufficient to capture
to high accuracy the entire multivariate time-series in $B$,
at least its components that are linearly predictable with the previous lag
of the time series from $C$ (this lag is the time series in $A$).

\begin{figure}
\centering
\includegraphics[width=\textwidth]{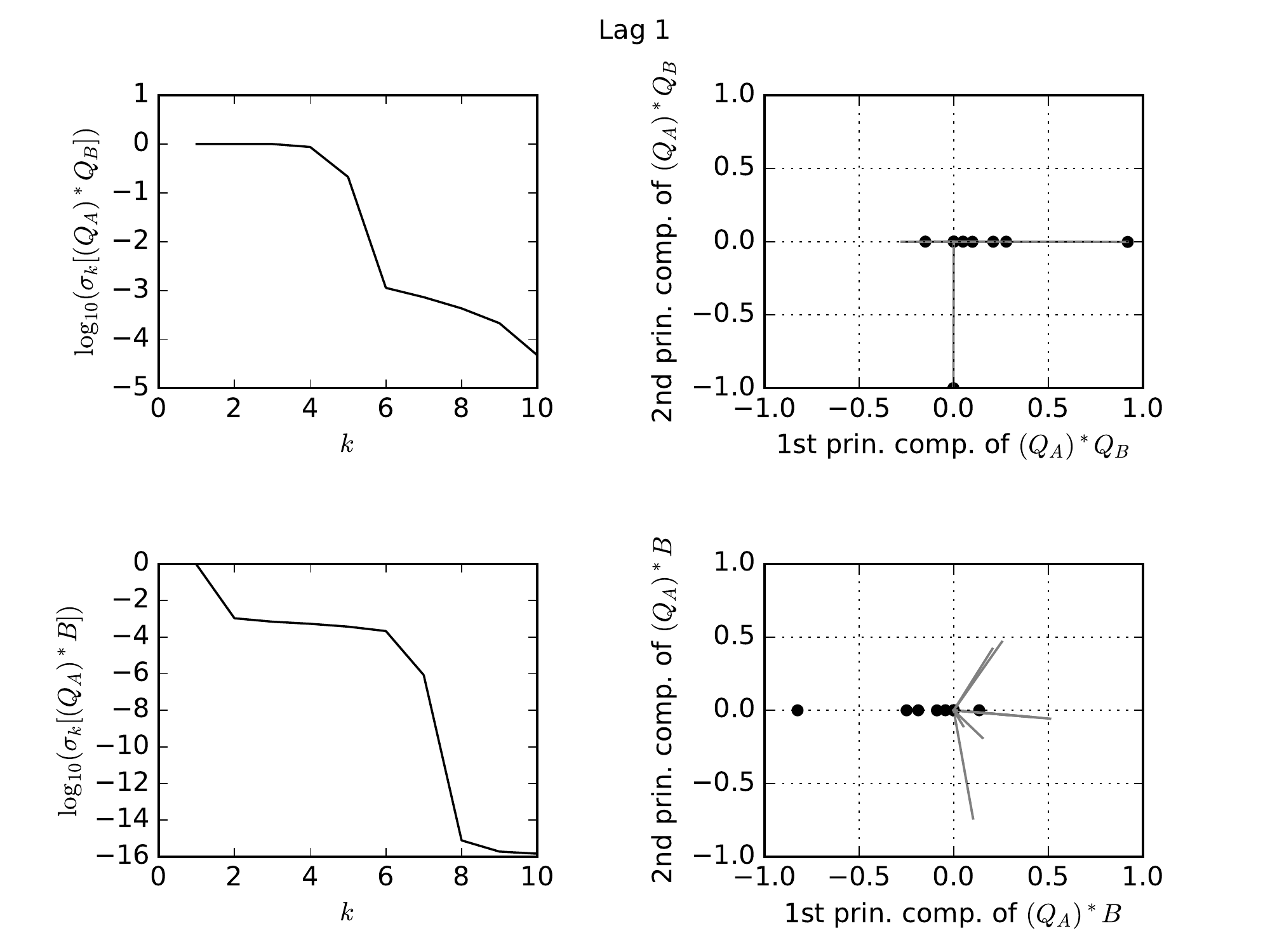}
\caption{Example from Subsection~\ref{synthseries}}
\label{synthplot}
\end{figure}

Figure~\ref{synthplot} displays the singular values
both for the matrix in the CCA between $A$ and $B$
and for the RAPCA of $B$ for $A$
(the former are in the top-left plot; the latter are in the bottom-left plot).
Regarding the biplots in the rightmost half of Figure~\ref{synthplot},
please consult~\cite{gabriel} or~\cite{gower-lubbe-le_roux}.
Figure~\ref{synthplot} shows that the spectral-norm accuracy
of the rank-4 RAPCA is similar to the excellent accuracy
of the corresponding RAID, whereas the spectral-norm accuracy
of the rank-4 CCA is very poor.

\subsection{Electricity loads}
\label{ld}

This subsection considers electricity meter readings for 370 clients
of a utility company from Portugal, with 140,256 readings per customer
in total; this data from~\cite{lichman} is available at
\url{http://archive.ics.uci.edu/ml/datasets/ElectricityLoadDiagrams20112014}
together with its complete detailed specifications.
We collect together the data into a 140,256 $\times$ 370 matrix $C$.
In this subsection, we rescale each column of $C$
so that its Euclidean norm becomes 1, thus ``equalizing'' clients.
For varying values of a lag $l$ (namely $l = 100$, 200, 300),
we let $A$ be the block of all rows of $C$ except the last $l$,
and let $B$ be the block of all rows of $C$ except the first $l$.
We then divide each entry of both $A$ and $B$ by the same value,
such that the spectral norm $\|B\|_2$ becomes 1.

\begin{figure}
\centering
\includegraphics[width=\textwidth]{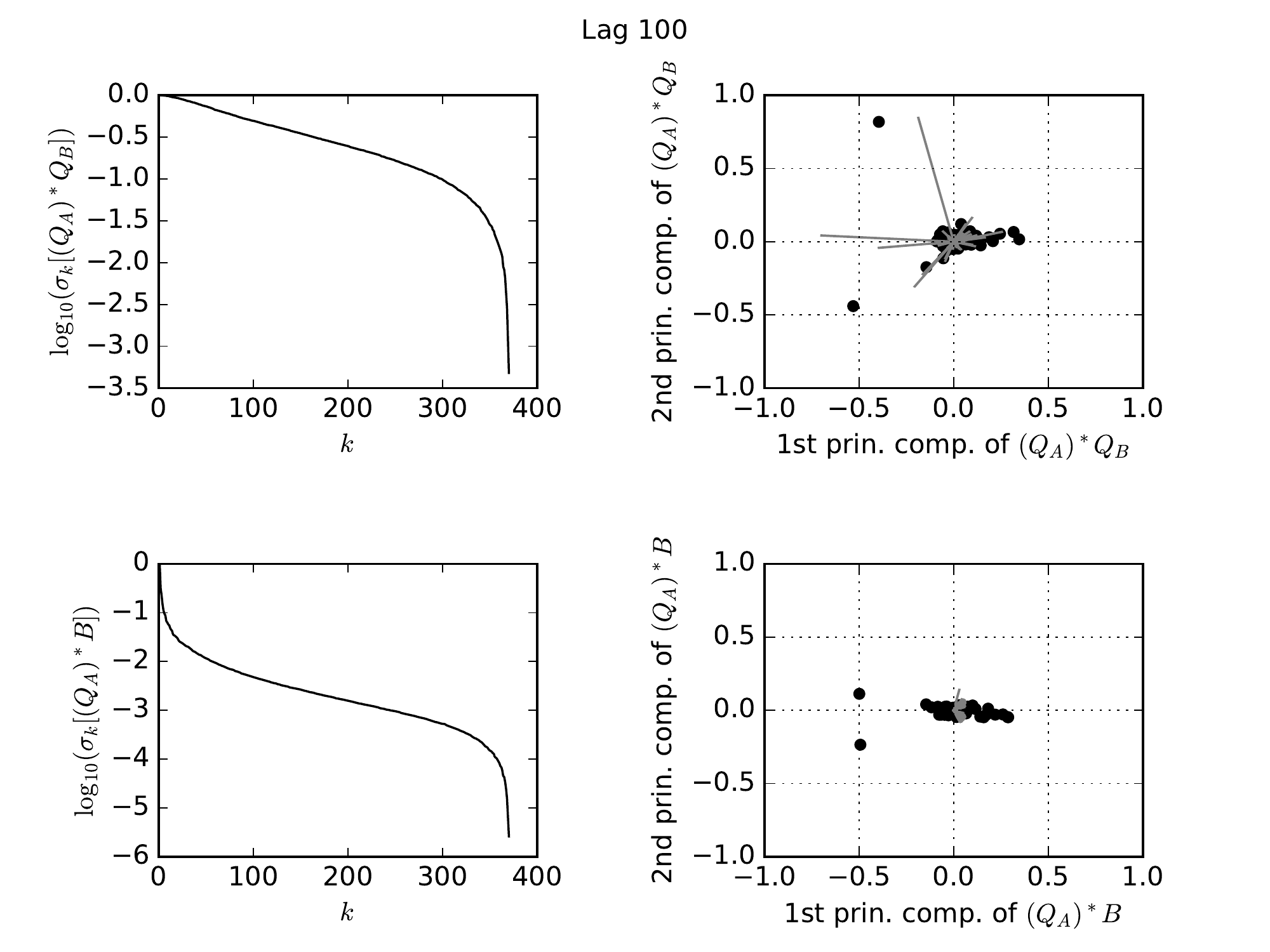}
\caption{Example from Subsection~\ref{ld} with $l = 100$}
\label{ld100}
\end{figure}

\begin{figure}
\centering
\includegraphics[width=\textwidth]{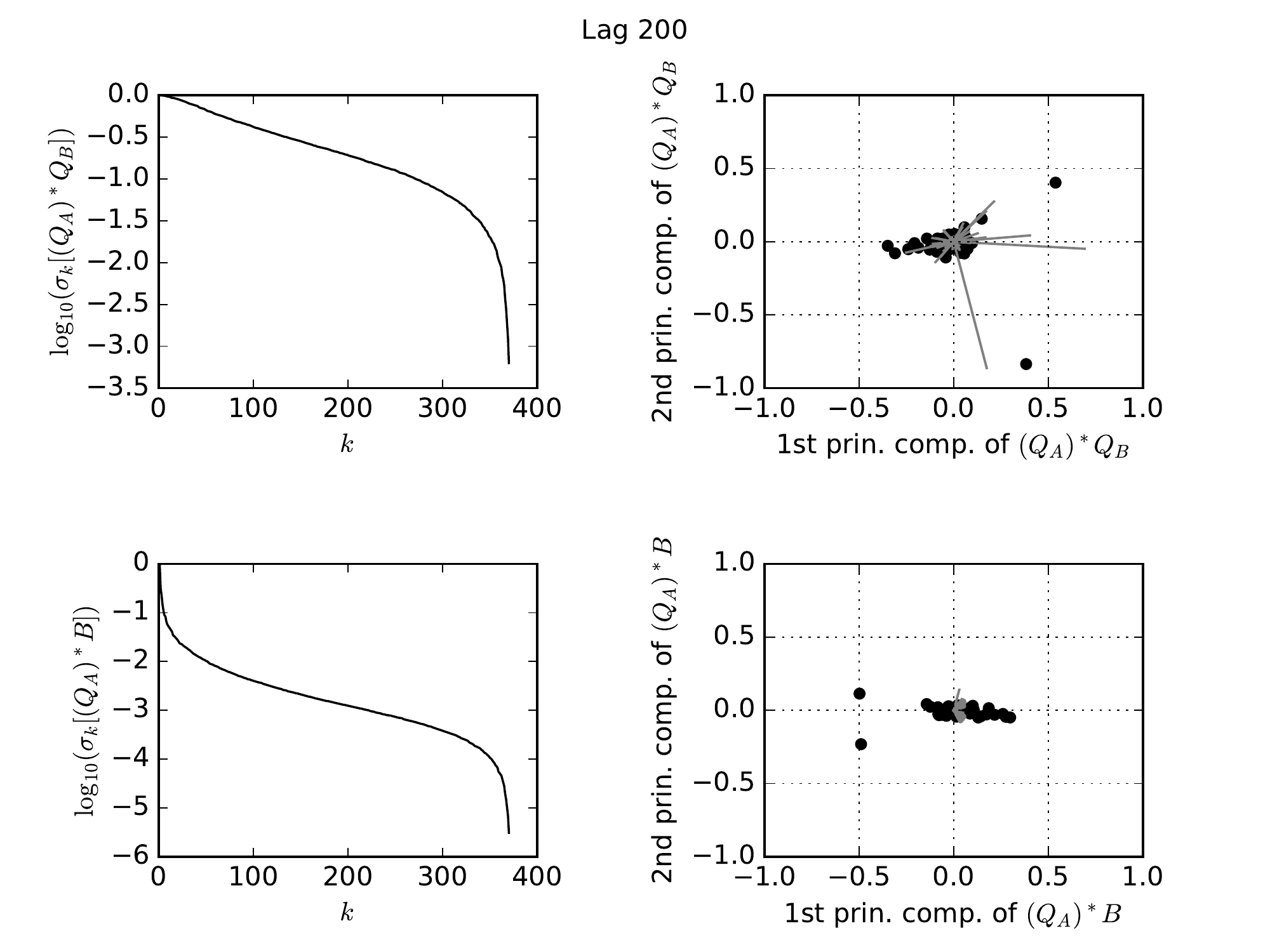}
\caption{Example from Subsection~\ref{ld} with $l = 200$}
\label{ld200}
\end{figure}

\begin{figure}
\centering
\includegraphics[width=\textwidth]{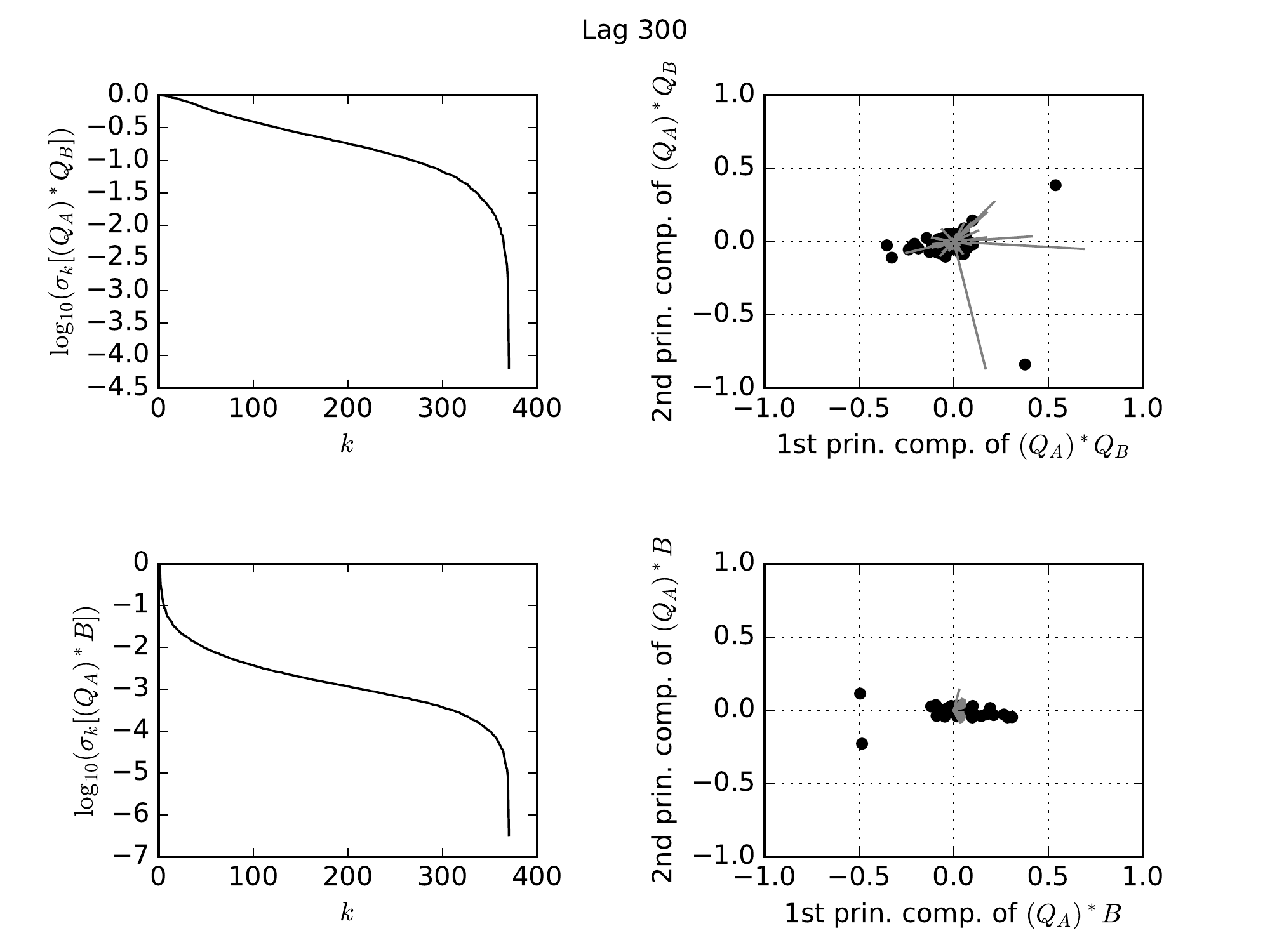}
\caption{Example from Subsection~\ref{ld} with $l = 300$}
\label{ld300}
\end{figure}

The ID of $B$ considered here selects 200 representative columns
from the originals; the RAID of $B$ for $A$ selects a different set of 200.
Table~\ref{ldtab} reports the spectral-norm accuracies attained.
Figures~\ref{ld100}--\ref{ld300} display the singular values
both for the matrix in the CCA between $A$ and $B$
and for the RAPCA of $B$ for $A$
(the former are in the top-left plot of each figure;
the latter are in the bottom-left plot of each figure).
Regarding the biplots in the rightmost halves
of Figures~\ref{ld100}--\ref{ld300},
please consult~\cite{gabriel} or~\cite{gower-lubbe-le_roux}.
Figures~\ref{ld100}--\ref{ld300} show that the spectral-norm accuracy
of the rank-200 RAPCA is similar to the high accuracy
of the corresponding RAID, whereas the spectral-norm accuracy
of the rank-200 CCA is two orders of magnitude worse.

\begin{table}
\caption{Example from Subsection~\ref{ld}}
\label{ldtab}
\begin{center}
{
\begin{tabular}{cccc}
$l$ & $\min_X \|AX-B\|_2$ & ID error & RAID error \\\hline
100 &                .075 &     .020 &      .0037 \\
200 &                .094 &     .020 &      .0030 \\
300 &                .098 &     .020 &      .0029 \\\hline
\end{tabular}
}
\end{center}
\end{table}

\subsection{Electricity loads transposed}
\label{ldt}

The present subsection considers the same data
as in the previous subsection, Subsection~\ref{ld}, but now
we let $B$ be the transpose of the block of the last 100,000 rows of $C$,
and let $A$ be the transpose of the block of the 300 rows just before $B$.
We then divide each entry of both $A$ and $B$ by the same number,
such that the spectral norm $\|B\|_2$ becomes 1.
The aim here is to select a small number, say 3, of the columns of $B$
that represent all 100,000 columns of $B$, or rather represent those components
which are linearly predictable with columns from $A$.
Thus, whereas the previous subsection selected representative clients,
with the clients' histories regressed against the lagged histories,
the present subsection selects representative times that the electricity meters
were read, with each time-slice of meter readings predicted
from the early readings collected together in $A$.
Transposing makes the columns refer to time-slices rather than clients
(rows then refer to clients).

As just mentioned, the ID of $B$ considered here selects 3 representative
columns from the originals; the RAID of $B$ for $A$ selects a different set
of 3.
The spectral norm of the difference between $B$ and its reconstruction
from the ID is 0.12.
The spectral-norm error of the RAID is 0.044.
(For reference, $\min_X \|AX-B\|_2 = 0.22$,
where $\|AX-B\|_2$ is the spectral norm of $AX-B$.)

\begin{figure}
\centering
\includegraphics[width=\textwidth]{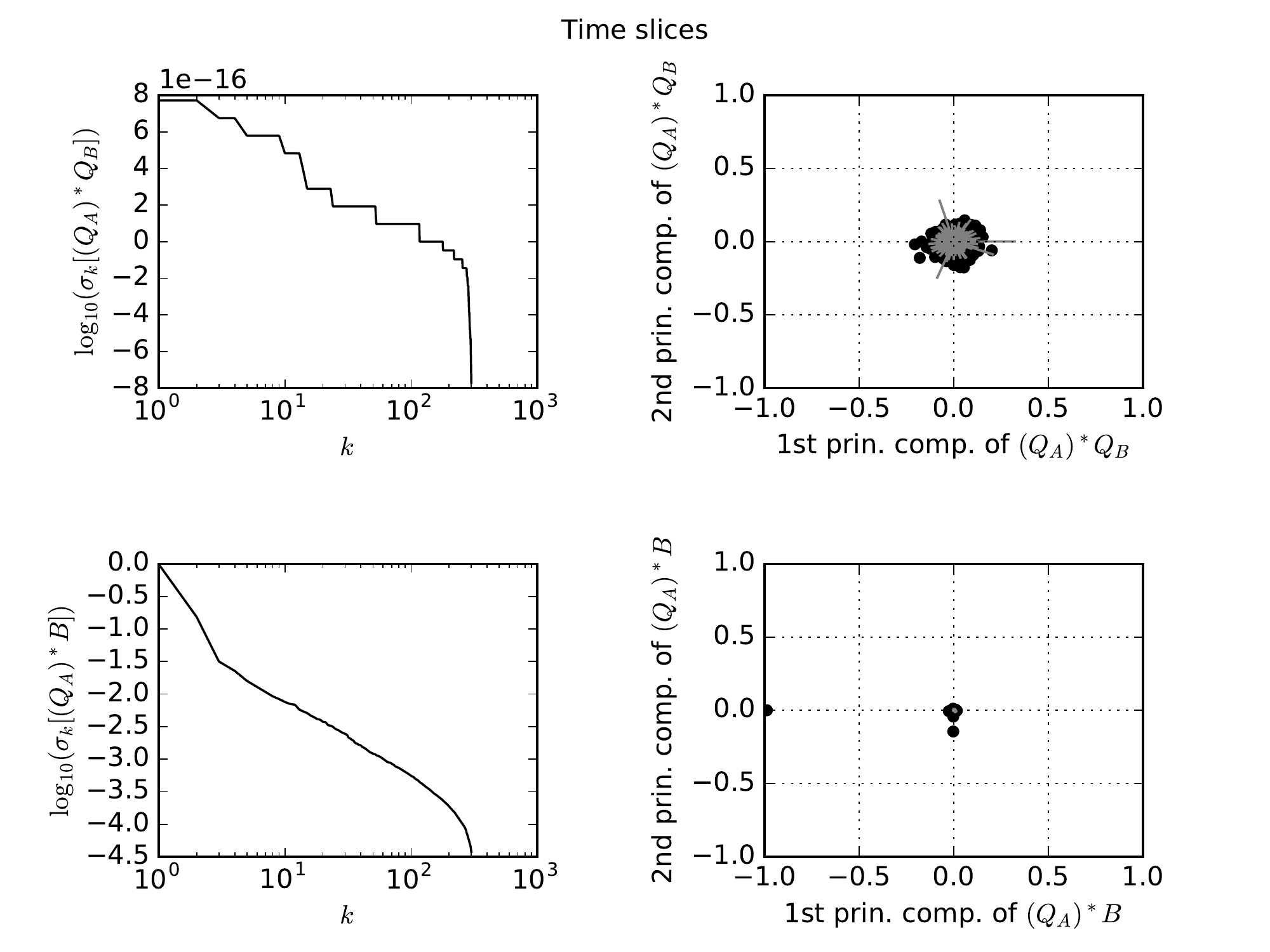}
\caption{Example from Subsection~\ref{ldt}}
\label{timeframes}
\end{figure}

Figure~\ref{timeframes} displays the singular values
both for the matrix in the CCA between $A$ and $B$
and for the RAPCA of $B$ for $A$
(the former are in the top-left plot; the latter are in the bottom-left plot).
Figure~\ref{timeframes} shows that the spectral-norm accuracy
of the rank-3 RAPCA is similar to the accuracy
of the corresponding RAID, whereas the CCA is vacuous (the logarithms
of all singular values in the top-left plot of Figure~\ref{timeframes}
are equal to 0 to nearly the machine precision of 0.22E--15).

\subsection{Motion capture}
\label{motion}

This subsection considers real-valued features derived
from motion-capture data of a person gesticulating;
for each of 1,743 successive instants, there are 50 real numbers
characterizing the gesticulator's motions and positions. This data
of~\cite{wagner-peres-madeo-lima-freitas} and~\cite{lichman} is available at
\url{http://archive.ics.uci.edu/ml/datasets/Gesture+Phase+Segmentation}
together with its complete detailed specifications.
We collect together the data into a 1,743 $\times$ 50 matrix $C$.
For varying values of a lag $l$ (namely $l = 20$, 40, 60),
we let $A$ be the block of all rows of $C$ except the last $l$,
and let $B$ be the block of all rows of $C$ except the first $l$.
We rescale each column of $A$ and each column of $B$
so that their Euclidean norms become 1.
We then divide each entry of both $A$ and $B$ by the same factor,
such that the spectral norm $\|B\|_2$ becomes 1.

\begin{figure}
\centering
\includegraphics[width=\textwidth]{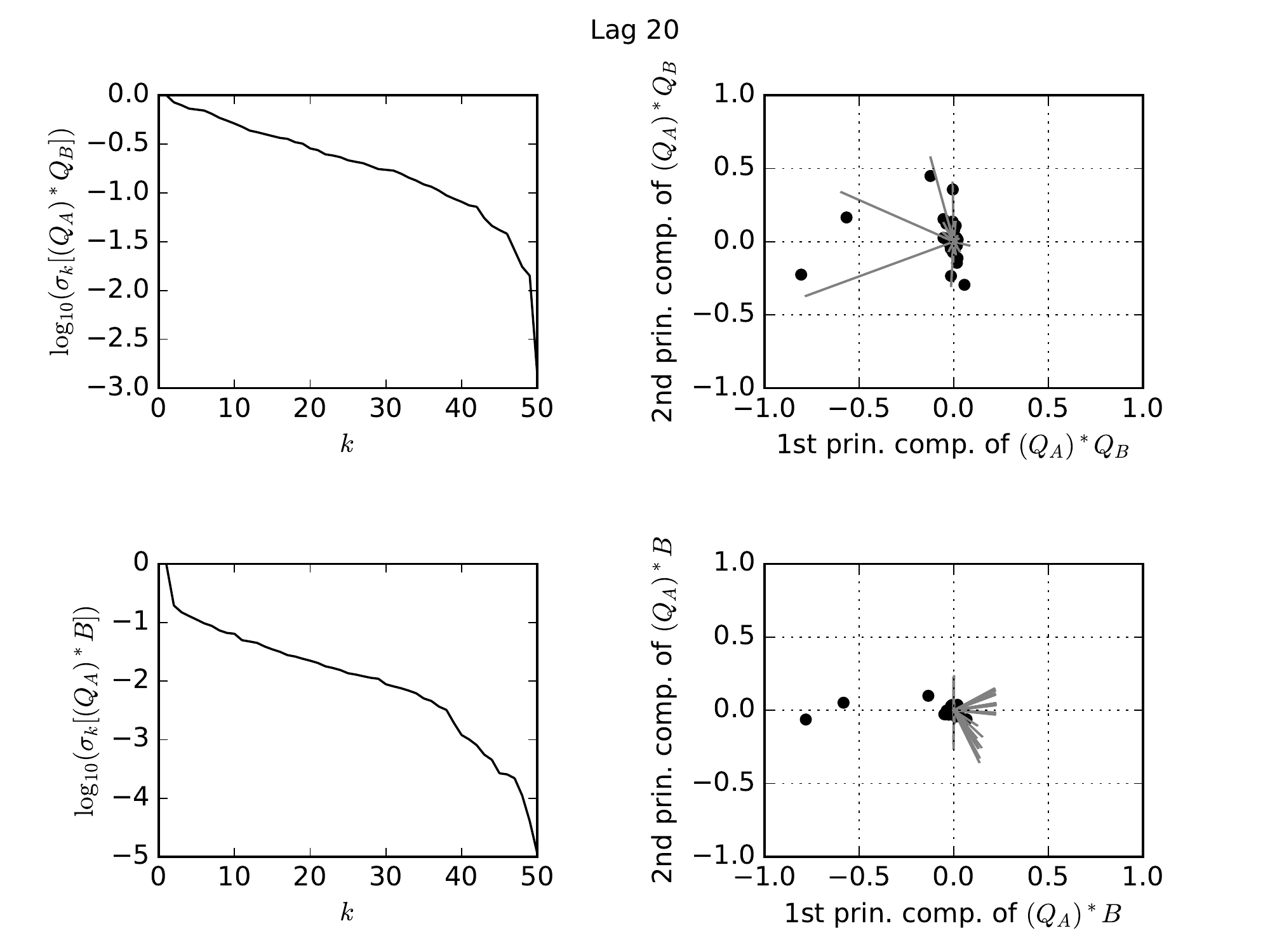}
\caption{Example from Subsection~\ref{motion} with $l = 20$}
\label{g20}
\end{figure}

\begin{figure}
\centering
\includegraphics[width=\textwidth]{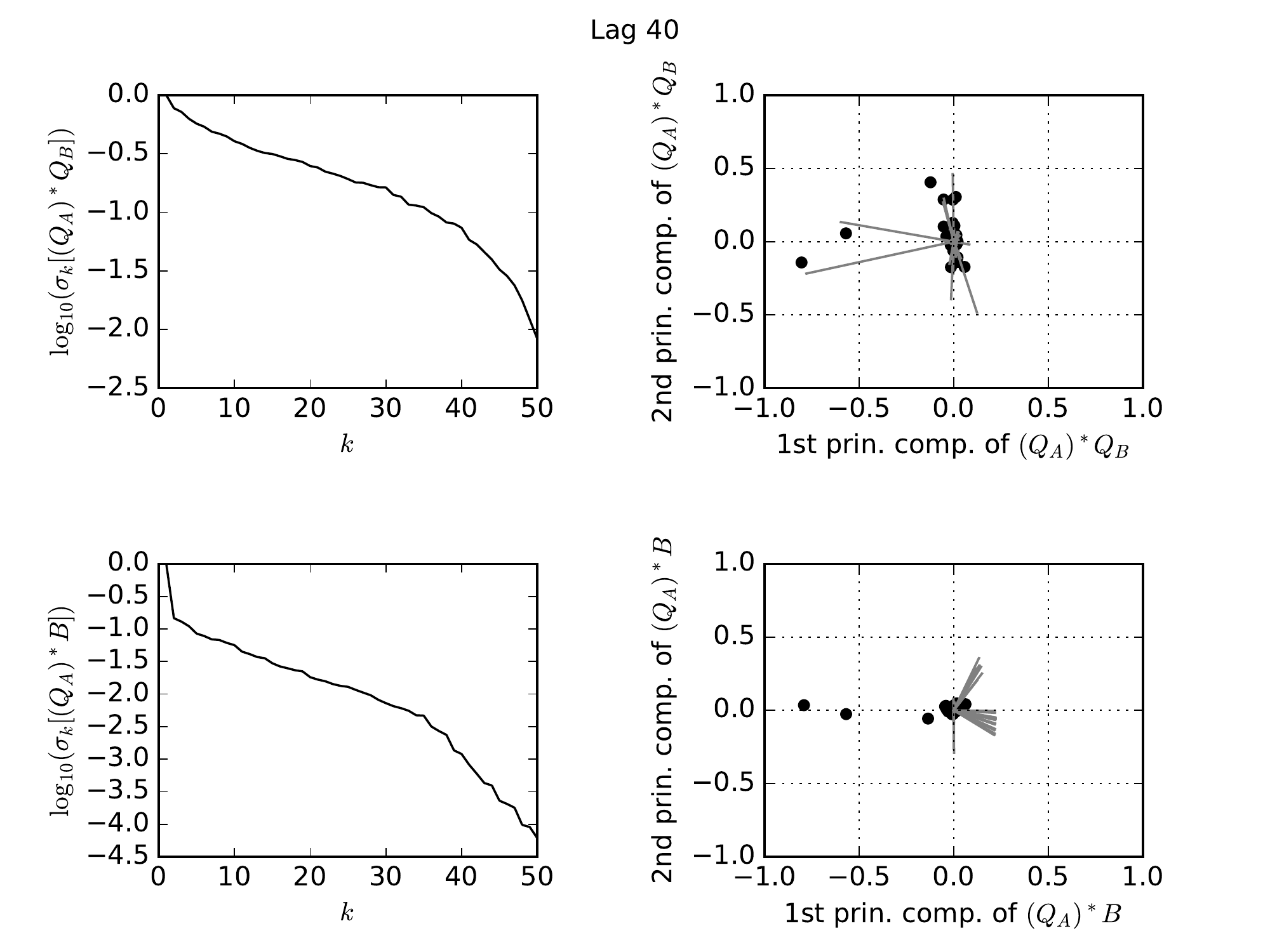}
\caption{Example from Subsection~\ref{motion} with $l = 40$}
\label{g40}
\end{figure}

\begin{figure}
\centering
\includegraphics[width=\textwidth]{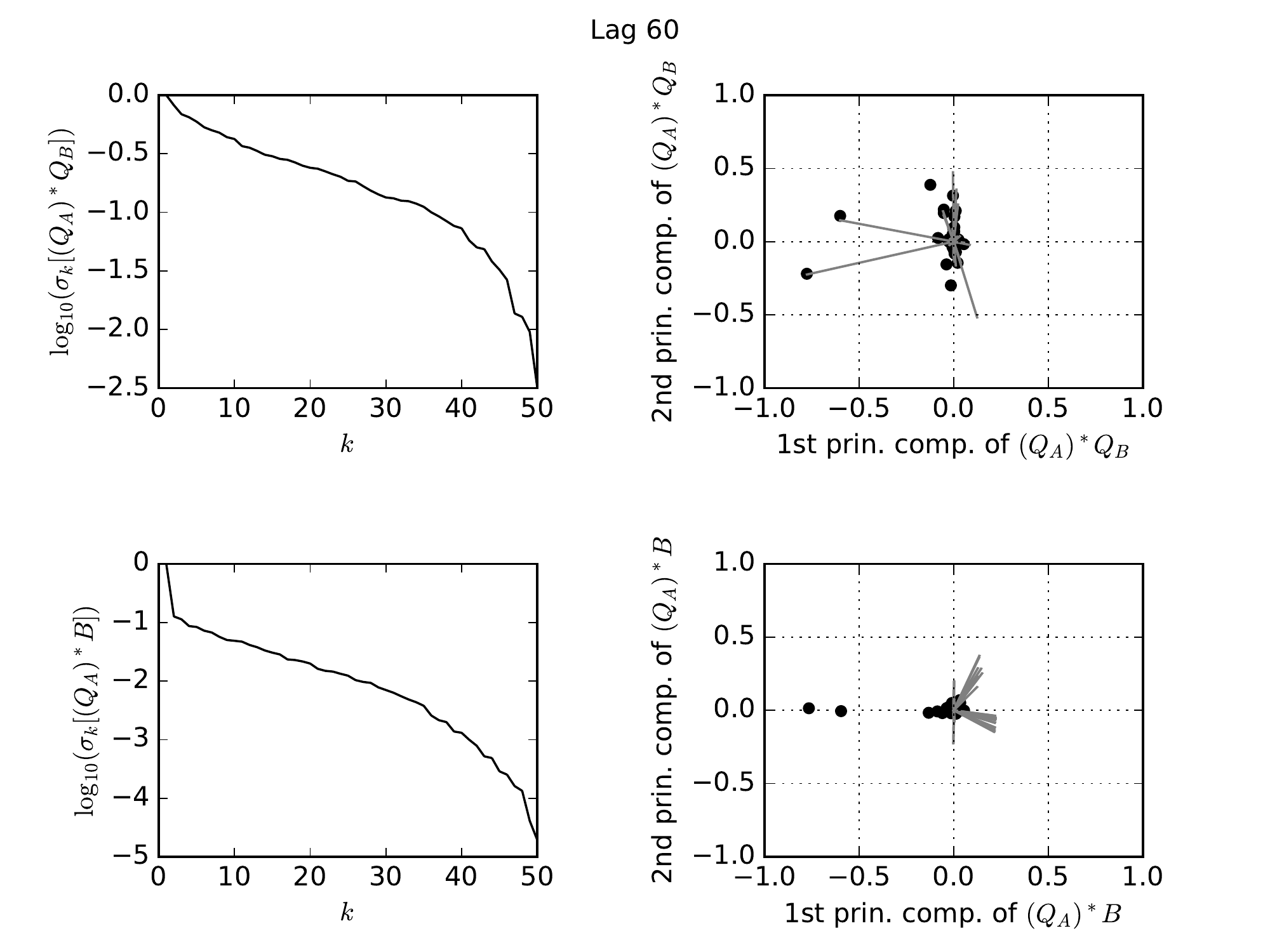}
\caption{Example from Subsection~\ref{motion} with $l = 60$}
\label{g60}
\end{figure}

Here, we consider an ID which selects 2 representative columns of $B$
and a RAID which selects a different 2.
Table~\ref{gtab} reports the spectral-norm accuracies attained.
Figures~\ref{g20}--\ref{g60} display the singular values
both for the matrix in the CCA between $A$ and $B$
and for the RAPCA of $B$ for $A$
(the former are in the top-left plot of each figure;
the latter are in the bottom-left plot of each figure).
Regarding the biplots in the rightmost halves
of Figures~\ref{g20}--\ref{g60},
please consult~\cite{gabriel} or~\cite{gower-lubbe-le_roux}.
Figures~\ref{g20}--\ref{g60} show that the spectral-norm accuracy
of the rank-2 RAPCA is similar to the accuracy of the corresponding RAID,
whereas the spectral-norm accuracy of the rank-2 CCA
is nearly the worst possible.

\begin{table}
\caption{Example from Subsection~\ref{motion}}
\label{gtab}
\begin{center}
{
\begin{tabular}{cccc}
$l$ & $\min_X \|AX-B\|_2$ & ID error & RAID error \\\hline
 20 &                 .41 &      .81 &        .16 \\
 40 &                 .41 &      .78 &        .15 \\
 60 &                 .42 &      .78 &        .13 \\\hline
\end{tabular}
}
\end{center}
\end{table}

\newpage

\bibliography{raid}
\bibliographystyle{siam}

\end{document}